# K-method of calculating the mutual influence of nodes in a directed weight complex networks

Andrei Snarskii[a,b], Dmyto Lande[b,a], Dmyto Manko[b*]

[a] *NTUU "Kyiv Polytechnic Institute", Kyiv, Ukraine*
[b] *Institute for Information Recording NAS of Ukraine, Ukraine*
**dmitriy.manko@gmail.com*

## Abstract

A new characteristic of paired nodes in a directed weight complex network is considered. A method (named as K-method) of the characteristics calculation for complex networks is proposed. The method is based on transforming the initial network with the subsequent application of the Kirchhoff rules. The scope of the method for sparse complex networks is proposed. The nodes of these complex networks are concepts of the real world, and the connections have a cause-effect character of the so-called "cognitive maps". Two new characteristics of concept nodes having a semantic interpretation are proposed, namely "pressure" and "influence" taking into account the influence of all nodes on each other.

## Highlights

- It is possible to use the simplified calculation mechanism, similar to one, is applicable in electrical engineering (the K-method).
- The proposed algorithm is devoid of the main problems and contradictions inherent in the widely used impulse method
- Two new weight characteristics of the concept nodes are proposed, namely pressure and influence as results of accounting for the network structure and "collective interaction".
- The results obtained with the help of the K-method are correlating well enough with the results obtained by the impulse method.

## 1. Introduction

The theory of complex networks emerged from the development of graph theory. First of all the theory is associated with the need to solve network problems (problems of dimension and complexity of computation, nonlinearity of the processes of time-dependent of the structure) in the case when analytical graph approaches can no longer be used. There are numbers of graph structures considered as complex networks, including the simplest ones, in which nodes are linked by means of connections without particular weights. Nodes and connections are assigned to certain numerical values, such that the connections are directed to even more complex ones.

Directed networks with weighted connections arise while describing a wide variety of tasks. For example, the Lancaster's law of combat proposed in 1916 describes the interaction of the two warring parties as a simple network of two nodes and two links between them [1]. Later on this model was generalized for the case of a large number of parties involved into a conflict [2]. There are few more generalizations of the Lancaster's model, for example, the model used for description of guerrilla warfare [3] requires complex networks approach. Another problem of this theory lies in controlling linear and nonlinear systems. An example of the problem is the task of controlling collective behavior; methods of the theory of complex networks are also used for as well. In this we consider networks with directed weight links [4]. We also consider the well-known problem of PageRank computing [5], [6] as well as HITS (Hyperlink induced topic search) index [5], [7], [8] as the ranks of complex networks (for example the Internet).

And, finally, cognitive maps must be mentioned as well [9], [10]. A cognitive map is a directed graph with given weights of connections [11]. According to given rules (which can be different for different statements of the problem), in calculating these network structures, it is necessary to find the weights of the nodes and the strength of the mutual (often indirect) influence of the nodes.

It is necessary to clearly understand that the definition, setting of the task in the analysis of the network structure depends not only on the structure of the network, but, and to a lesser extent, on "rules of the game" of this network. Steady weights of nodes/links are calculated with those rules (or algorithms). Practically, these algorithms are completely different for different tasks.

The research of the mentioned (and many other) problems within the framework of the theory of complex networks can be conditionally divided into two parts. The first of which consists in a formal investigation of a given network, for example for establishing conditions for the existence of a solution (fixed point), its stability, etc [10], [6].

The second part of the problem is less formalized and consists of three parts, each of which is a significant problem itself. Thus, the first part is the construction of "rules of the game" - an algorithm that corresponds to the task within the task, allows calculating the weights of nodes in the PageRank task. And thirdly, the interpretation of the obtained results, namely the weights of nodes or links. The first and third problems relate to the understanding of the real world's subject areas, the rules of the functioning of the network models and their interpretations are completely dependent on experts in given areas.

After the network is defined, the initial values of the weights of the links and/or nodes are defined and “rules of the game" are obtained, the task is formalized. And the only thing remains is to find the final values of the weight values of the nodes (or links), namely the fixed point of the equations (in the case it exists) describing their changes. For such a formalized problem, the conditions make possible the fixed point existence are stable and should be defined.

## 2. Problem statement

In this part we considering processes, when some nodes, concepts of which correspond to the real world can influence others by changing their numerical values. The magnitudes of the influence are determined by preset weights of directed bonds, which can be either positive or negative. Next, we use the terminology of the theory of cognitive maps, naming the concept nodes as “concepts”.

Let's consider several problems related to the mutual influence of concept nodes on such graphs. The simplest problem is shown in Fig. 1a, having two concept nodes in the network, with the first node affecting the second node and the second node affecting the first node. The numerical value of the magnitude of the influence is denoted by $\varepsilon$. If the value of the 1st concept 1 is given and the magnitude of its influence on concept 2, then it is natural to assume as follows: $\varphi_2 = \varphi_1 + \varepsilon$, which is

$$\varphi_2 - \varphi_1 = \varepsilon . \quad (1)$$

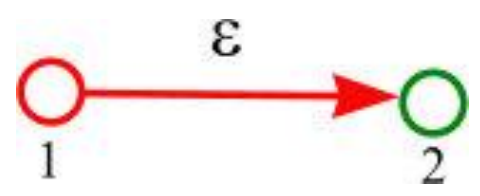

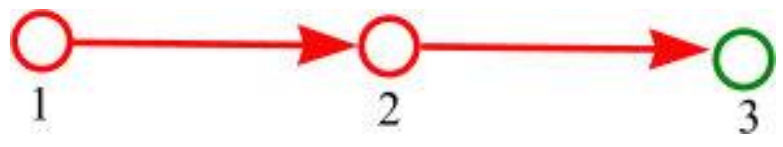

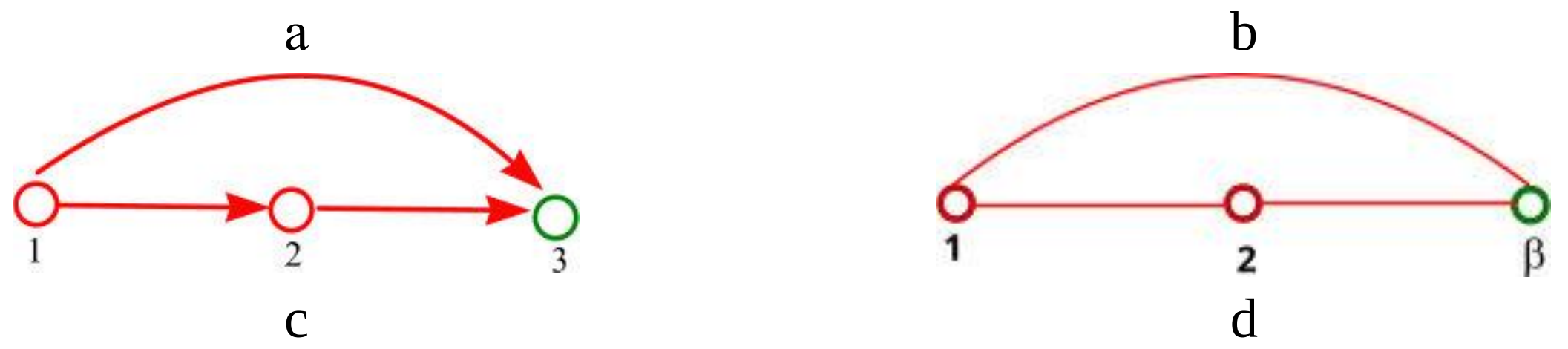

Fig. 1 - Examples of the simplest influence graphs

A few more simple cases of the networks are shown in Fig. 1b, 1c, 1d. In the network depicted (Fig. 1b) there is no direct influence of concept 1 on concept 3, thereby $\boldsymbol{\varepsilon}_{13} = 0$. However, there is an indirect effect of the 1-factor on the 3-factor by the concept 2 ($\boldsymbol{\varepsilon}_{12} \neq 0$, $\boldsymbol{\varepsilon}_{23} \neq 0$). The task is to determine the full impact of one concept on the other in the absence of a direct link. Thus it is needed to determine influence, which includes not only direct (in the case of their existence), but also all indirect influences. It is supposed that if one can find a path (via an oriented graph) from one concept (for example $\alpha$), to another (for example $\beta$), then there is an influence (the indirect one) of the concept $\alpha$ on the concept $\beta$. Let's denote the total, influence, covering influence over all possible paths as $K_{\alpha\beta}$. To put this another way, having a matrix with elements it is necessary to find the corresponding *K*-matrix, with elements $K_{ij}$, thereby describe the function $\boldsymbol{K} = \boldsymbol{K}(\boldsymbol{\varepsilon})$ . Elements of the matrix of $\boldsymbol{\varepsilon}$ define the direct influences between the concepts. Elements $K_{\alpha\beta}$ of the matrix $\boldsymbol{K}$ set all the influences, including indirect one. And, surely, in the general case inequality comes true: $K_{\alpha\beta} \neq \varepsilon_{\alpha\beta}$ (for example, for the cognitive map shown in Fig. 1b $\varepsilon_{13} = 0$, while $K_{13} \neq 0$). Accordingly, the networks that correspond to the matrices of contiguity $\varepsilon_{ik}$ and the matrix $\boldsymbol{K} = \boldsymbol{K}(\boldsymbol{\varepsilon})$ are different networks.

Let's list the requirements that the function $\boldsymbol{K}(\boldsymbol{\varepsilon})$ should satisfy:

1. Linearity, $\boldsymbol{K}(\boldsymbol{c} \cdot \boldsymbol{\varepsilon}) = \boldsymbol{c} \cdot \boldsymbol{K}(\boldsymbol{\varepsilon})$, $c$ =constant. In particular, it does not matter in which units the influence $\varepsilon_{\alpha\beta}$ is given.

2. For any given matrix $\boldsymbol{\varepsilon}$ there can be found and a $\boldsymbol{K}$ matrix.

3. For finite values of $\boldsymbol{\varepsilon}$ matrix elements, the matrix values $\boldsymbol{K}$ must be finite.

4. The result of the calculation must be mathematically stable. Small deviations of $\boldsymbol{\varepsilon}$ should lead to small deviations of $\boldsymbol{K}$. This means that if $\boldsymbol{K}(\boldsymbol{\varepsilon} + \delta\boldsymbol{\varepsilon}) = \boldsymbol{K}(\boldsymbol{\varepsilon}) + \delta\boldsymbol{K}(\boldsymbol{\varepsilon})$ and $\delta\boldsymbol{\varepsilon} / \boldsymbol{\varepsilon} \ll 1$, then $\delta\boldsymbol{K} / \boldsymbol{K} \ll 1$.

## 3. Method of calculating the *K*-matrix

The main idea of the method consists in the analogy of calculating the influence in an electrical circuit's network. Considering this analogy, it is assumed that the direct power $\varepsilon_{ik}$ of the $i$th concept's influence on the concept $k$ is an analog of the electromotive force (EMF) in the electric circuit $ik$. In this case $\varphi_k - \varphi_i$ is an analogue of the difference of electric potentials between the corresponding nodes. Then $K_{\alpha\beta}$ is the potential difference $\varphi_\alpha - \varphi_\beta$ with the specified EMF in a electrical circuit. The resistance of a subcircuit of the network is assumed to be equal to 1. Such an analogy leads to the fact that the calculation $\boldsymbol{K} = \boldsymbol{K}(\boldsymbol{\varepsilon})$ must be equivalent to solving the Kirchhoff equations for linear electrical circuit.

Unfortunately, in the proposed case, such a simple and transparent analogy cannot be realized. The point is that the considered networks are digraphs, which, from the point of view of the proposed "electrical" analogy, are electrical circuit containing diodes that transmit currents and charges in only one direction.

This means that for the network shown in Fig. 1a in accordance to formula (1) relation $\varphi_2 - \varphi_1 = \varepsilon$ come true, in the case of $\varepsilon > 0$, $\varphi_2 - \varphi_1 = 0$. Thus, a sign $\varepsilon$ specifies "positivity" or "negativity" of an influence may contradict a given direction of influence, which does not allow taking into account negative influences.

In order to avoid this contradiction, the below described method is proposed. The method allows calculating the measured influence of one concept on another (Fig. 2) (for example to determine the influence, of a concept $\alpha$ on a concept $\beta$ in a given network (Fig. 2a)):

1. There are all possible paths without loops from node $\alpha$ to node $\beta$ (Fig. 2b). At the same time, in each of the path, a link can enter into any node no more than once.

2. All the paths obtained are connected to a single network in the way that all paths start from one node $\alpha$ and terminate $\beta$ at a node. 2c.

3. In the obtained graph (which, strictly speaking, is not a subgraph of the network under consideration), the direction of the links is removed. The resulting graph is called the $K_{\alpha\beta}$-graph. We note at once that in this case different graphs correspond to different pairs of indices.

4. The numerical value of the concept's $\alpha$ influence on the concept $\beta$, which is denoted by the same symbol $K_{\alpha\beta}$, is calculated according to Kirchhoff's rules (now for the network obtained in step 3, which does not contain non-linear elements, thus the solution always exists and is unique)

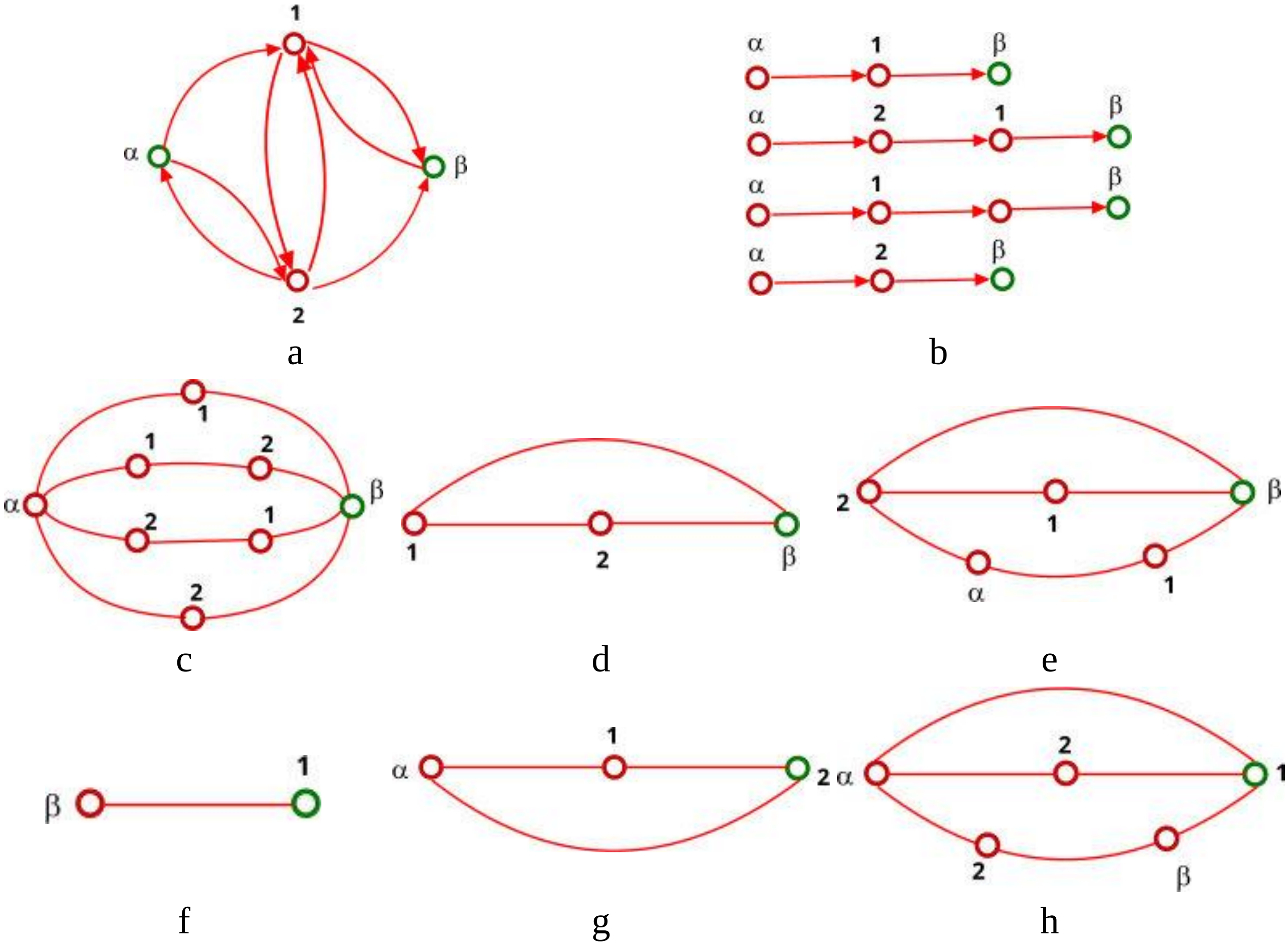

Fig. 2 - Example of calculating the influence of concepts

Fig. 3. demonstrates a network for the general case (for an arbitrary network) obtained for selected nodes $\alpha$ and $\beta$ that are connected by $M$ parallel paths, each of which consists of $N_m$ consecutive links $(m = 1, 2, ..., M)$. Accordingly $n^{(m)}$, this is the $n$ th bond in the $m$ th path, containing the EMF, equal to $\varepsilon_m^{n^{(m)}}$.

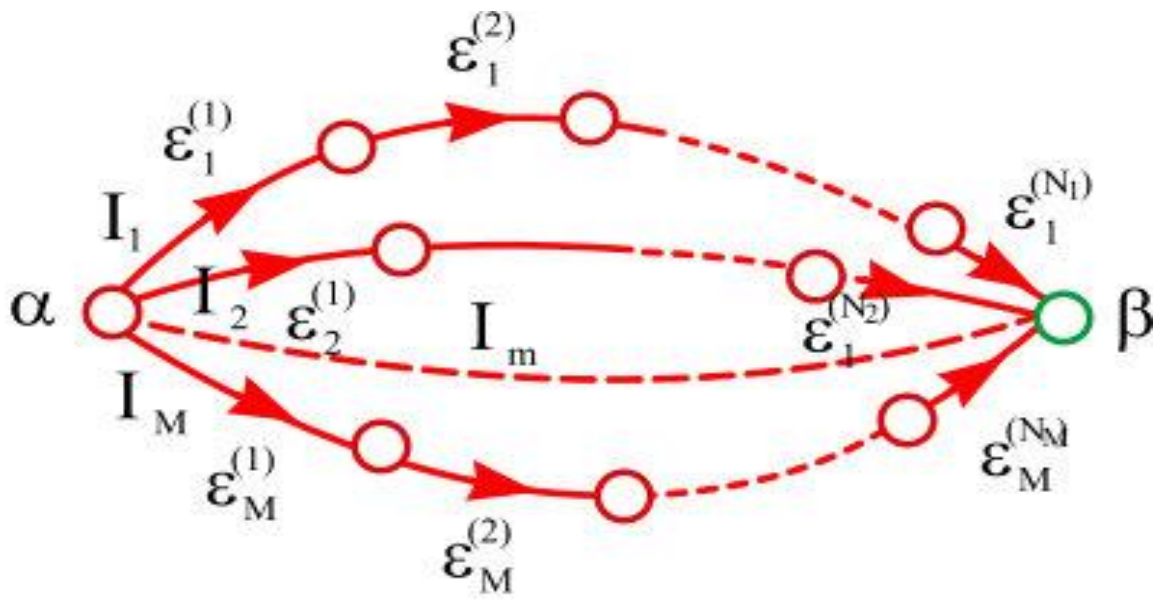

Fig. 3 - General view of the electrical circuit for a graph consisting of parallel paths.

The desired value of the concept's influence $\alpha$ on the concept $\beta$, which is defined as the electric potential difference $K_{\alpha\beta} = \varphi_\beta - \varphi_\alpha$, is calculated in accordance with Kirchhoff's rules:

$$K_{\alpha\beta} = \frac{\sum_{m=1}^{M} (\varepsilon_m / N_m)}{\sum_{m=1}^{M} (1/N_m)}, \quad \varepsilon_m = \sum_{n^{(m)}=1}^{N_m} \varepsilon_m^{(k)}, \tag{2}$$

where $\boldsymbol{\varepsilon}_m$ is the total EMF in the $m$-th path, $N_m$-the number of connections along this path.

## 4. An example of matrix $\boldsymbol{K}$ calculation

Let's consider the network shown in Fig. 2a. The case is chosen when, at each connection, the force of influence is positive and identical, which for simplicity is taken equal to unity. In this case the matrices have the form

$$\boldsymbol{\varepsilon} = \begin{array}{c|cccc} & \alpha & 1 & 2 & \beta \\ \hline \alpha & 0 & 1 & 1 & 0 \\ 1 & 0 & 0 & 1 & 1 \\ 2 & 1 & 1 & 0 & 1 \\ \beta & 0 & 1 & 0 & 0 \end{array}, \quad K = \begin{array}{c|cccc} & \alpha & 1 & 2 & \beta \\ \hline \alpha & 0 & 1.6 & 1.3 & 2.4 \\ 1 & 2 & 0 & 1 & 1.3 \\ 2 & 1 & 1.5 & 0 & 1.64 \\ \beta & 3 & 1 & 2 & 0 \end{array}. \tag{3}$$

It should be noted, that no network $\boldsymbol{K}$ corresponds to the matrix, but each element of the matrix has its own network. For example, the element $K_{1\beta}$ of the matrix $\boldsymbol{K}$ corresponds to the network shown in Fig. 2d.

The matrix $\boldsymbol{K}$ obtained in (3) allows ranking paired nodes according to the magnitude of the effect of one node on another, taking into account all mediated influences. According to (3), the maximum value of the influence corresponds to the pair of $\alpha \to \beta$ and the minimum value to the pair of $1 \to 2$, $1 \to \alpha$ and $\beta \to 1$.

This, ranking of paired nodes is of independent interest in the analysis of networks. However, using the paired ranking, one can additionally get two types of nodes ranking, which can be called "pressure" and "influence."

## 5. Nods ranking. Pressure and influence

In accordance with (3), the K-method makes possible to calculate the mutual influence of one concept (α) on another concept (β), taking into account the influence of all nodes on each other. We can calculate such characteristics of the cognitive map as "paired influences".

In addition to this "paired influence" based on the $\boldsymbol{K}$-matrix, we can introduce the notion of "collective interaction" and calculate it. Let us consider two above mentioned characteristics, namely "pressure" - ψ and "influence" - ν, (Fig. 4).

It should be noted that the arrows on the relationships on the diagrams do not correspond to the influences on the cognitive map, however, they mean the existence of the corresponding component of $\boldsymbol{K}$-matrix.

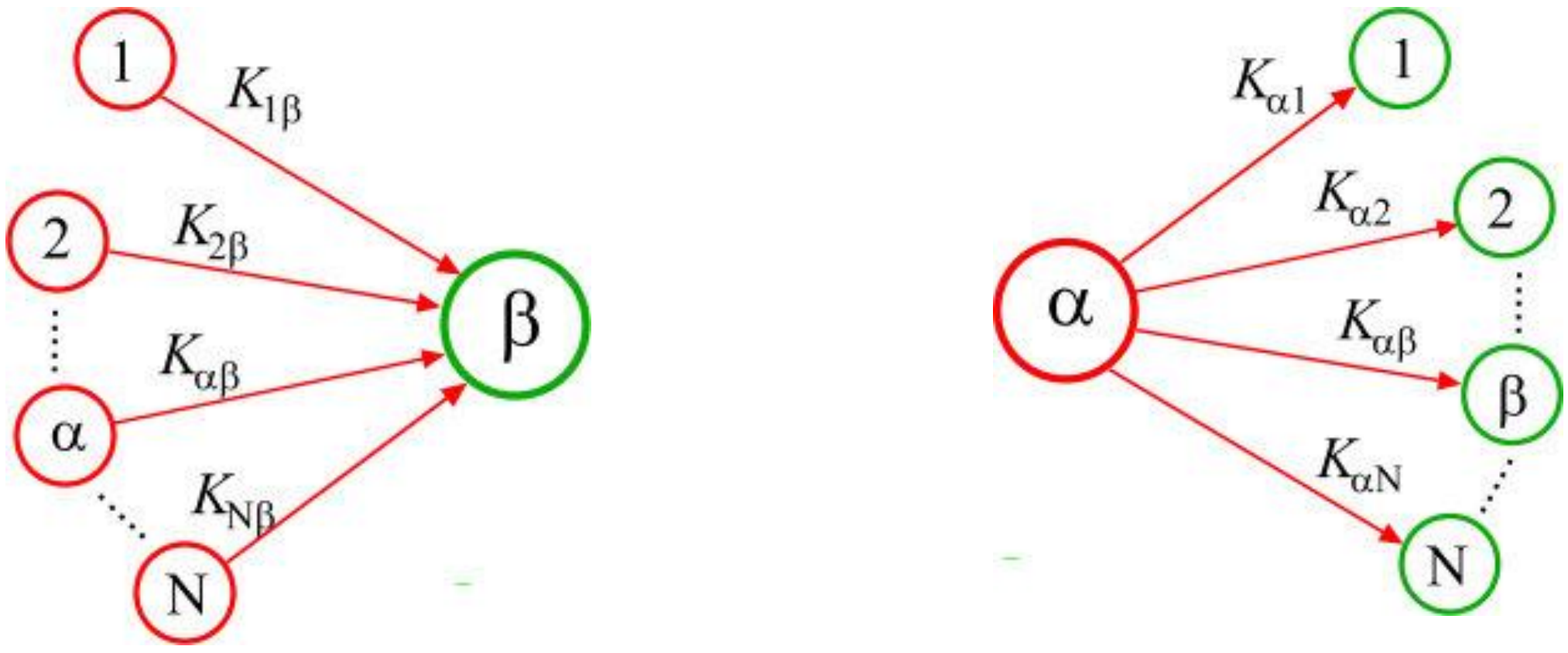

a b

Fig. 4 - Characteristics of cognitive maps – pressure, ψ (a) and influence, ν (b)

The components of the vectors ψ and ν are calculated on the basis of the matrix as follows:

$$\psi_\beta = \sum_\alpha K_{\alpha\beta}, \quad \nu_\alpha = \sum_\beta K_{\alpha\beta}. \tag{4}$$

Since these cognitive maps do not take into account the influence of the node on itself, the elements with α = β do not enter into expressions (4).

According to his definition (4), pressure describes the total impact of all other concepts on concept β, and influence $\nu_\alpha$ is the sum of the effects of concept α on all other concepts.

In addition to the definition (4), the amplitude pressure $a\psi_\beta$ and amplitude influence $a\nu_\alpha$ are introduced, their absolute values can be calculated as fillows:

$$a\psi_\beta = \sum_\alpha |K_{\alpha\beta}|, \quad a\nu_\alpha = \sum_\beta |K_{\alpha\beta}|. \tag{5}$$

Thus, to calculate, for example $a\nu_{\alpha}$, the sign of the influence of the concept α on concept β is not important, only its value is taken into account.

Table 1. Values of ψ and ν for the network shown in Fig. 2 arranged in accordance with the ranks

| | | | |
|---|---|---|---|
| $\psi_{\alpha} = 5.3$ | $\psi_1 = 4.3$ | $\psi_2 = 4.14$ | $\psi_{\beta} = 6$ |
| $\nu_{\beta} = 5.34$ | $\nu_2 = 4.3$ | $\nu_1 = 4.1$ | $\nu_{\alpha} = 6$ |

Let's consider an example of the cognitive map given in [8], connected with the problem of fuel consumption and maintenance of air pollution free.

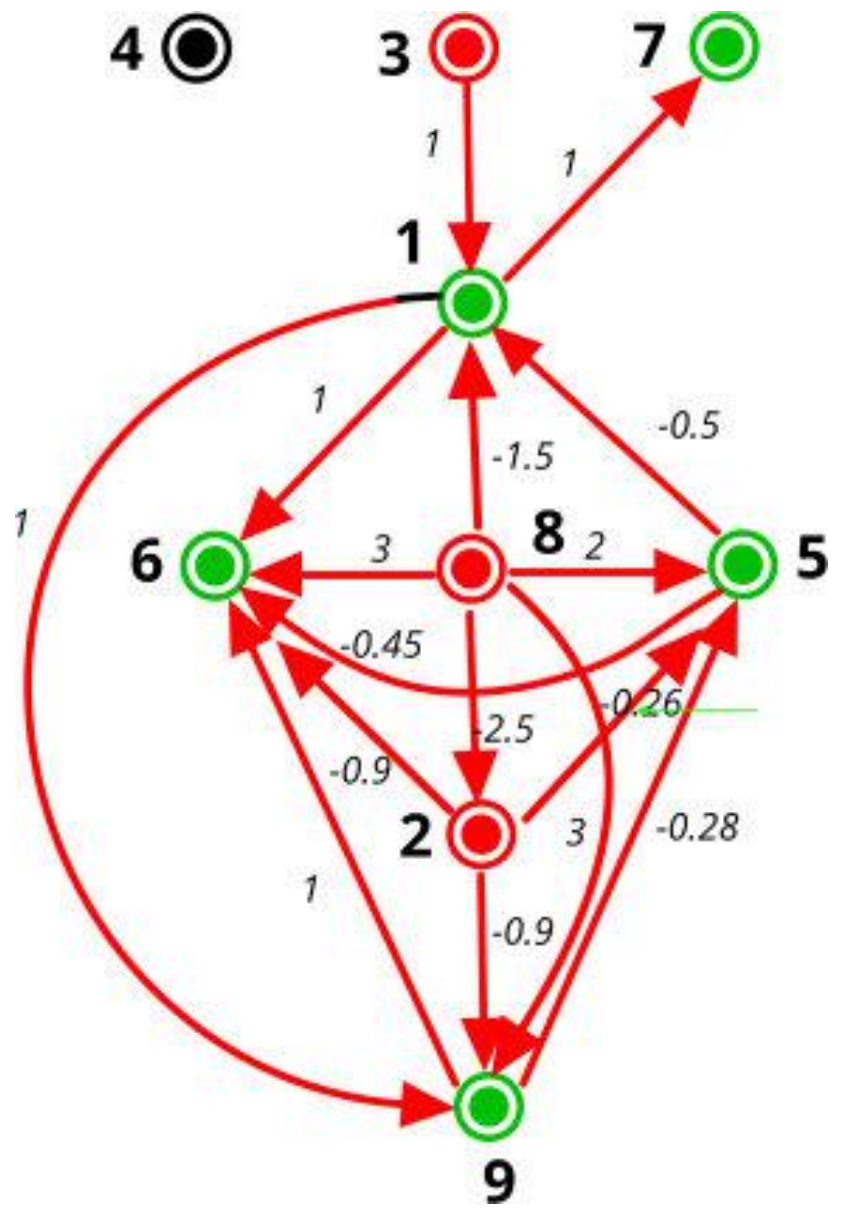

Fig. 5. The cognitive map for analyzing the problem of fuel consumption and maintaining air purity in the city of San Diego, California [8]: 1 - the length of the trip, 2 - fuel saving, miles per gallon; 3 - population; 4 - the cost of the car; 5 - cost of the trip; 6 - volume of emissions of exhaust gases into the atmosphere; 7 - accidents; 8 -average delay; 9 - fuel consumption.

The adjacency matrix $W$ of the cognitive map is shown in Fig. 5 given as follows:

$$W = \begin{pmatrix} 0 & 0 & 0 & 0 & 0 & 1 & 1 & 0 & 1 \\ 0 & 0 & 0 & 0 & -0.26 & -0.9 & 0 & 0 & -0.9 \\ 1 & 0 & 0 & 0 & 0 & 0 & 0 & 0 & 0 \\ 0 & 0 & 0 & 0 & 0 & 0 & 0 & 0 & 0 \\ -0.5 & 0 & 0 & 0 & 0 & -0.45 & 0 & 0 & 0 \\ 0 & 0 & 0 & 0 & 0 & 0 & 0 & 0 & 0 \\ 0 & 0 & 0 & 0 & 0 & 0 & 0 & 0 & 0 \\ -1.5 & -2.5 & 0 & 0 & 2 & 3 & 0 & 0 & 3 \\ 0 & 0 & 0 & 0 & -0.28 & 1 & 0 & 0 & 0 \end{pmatrix}, \qquad (6)$$

The K-matrix of the cognitive map shown in Fig. 5 given as follows:

$$K = \begin{pmatrix} 0 & 0 & 0 & 0 & 0.72 & 1.33 & 1 & 0 & 1 \\ -1.13 & 0 & 0 & 0 & -0.57 & -0.27 & -0.15 & 0 & -0.62 \\ 1 & 0 & 0 & 0 & 1.72 & 2.40 & 2 & 0 & 2 \\ 0 & 0 & 0 & 0 & 0 & 0 & 0 & 0 & 0 \\ -0.65 & 0 & 0 & 0 & -0.45 & 0.73 & 0.32 & 0 & 0.32 \\ 0 & 0 & 0 & 0 & 0 & 0 & 0 & 0 & 0 \\ 0 & 0 & 0 & 0 & 0 & 0 & 0 & 0 & 0 \\ -0.89 & -2.5 & 0 & 0 & 0.18 & 0.75 & 0.12 & 0 & 0.51 \\ -0.78 & 0 & 0 & 0 & -0.28 & 0.80 & 0.22 & 0 & 0 \end{pmatrix} \qquad (7)$$

Ranking by the magnitude of the elements of the matrix K is:

$$K_{36} > K_{37} = K_{39} > K_{35} > ... > K_{81} > K_{21} \qquad (8)$$

According to (7), the greatest influence on the amount of emissions to the atmosphere, taking into account all other influences, is provided by the number of people.

The result of calculating "pressure", $\psi$ - and "influence", $\nu$ according to the theory of the K-matrix (6) is shown in Fig. 6.

(a)

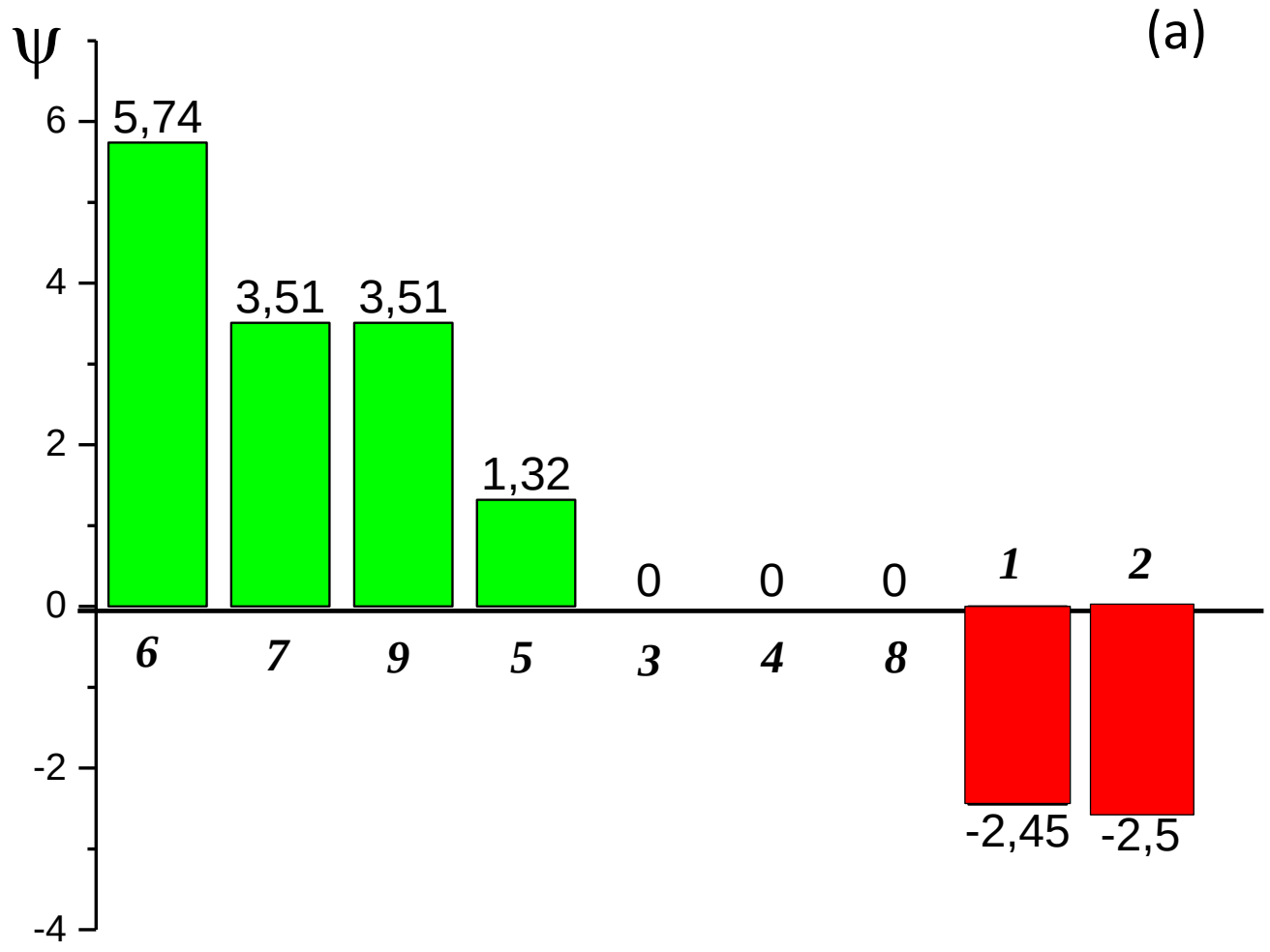

(b)

$$\psi = \begin{pmatrix} -2.45 \\ -2.5 \\ 0 \\ 0 \\ 1.32 \\ 5.74 \\ 3.51 \\ 0 \\ 3.21 \end{pmatrix}$$

(c)

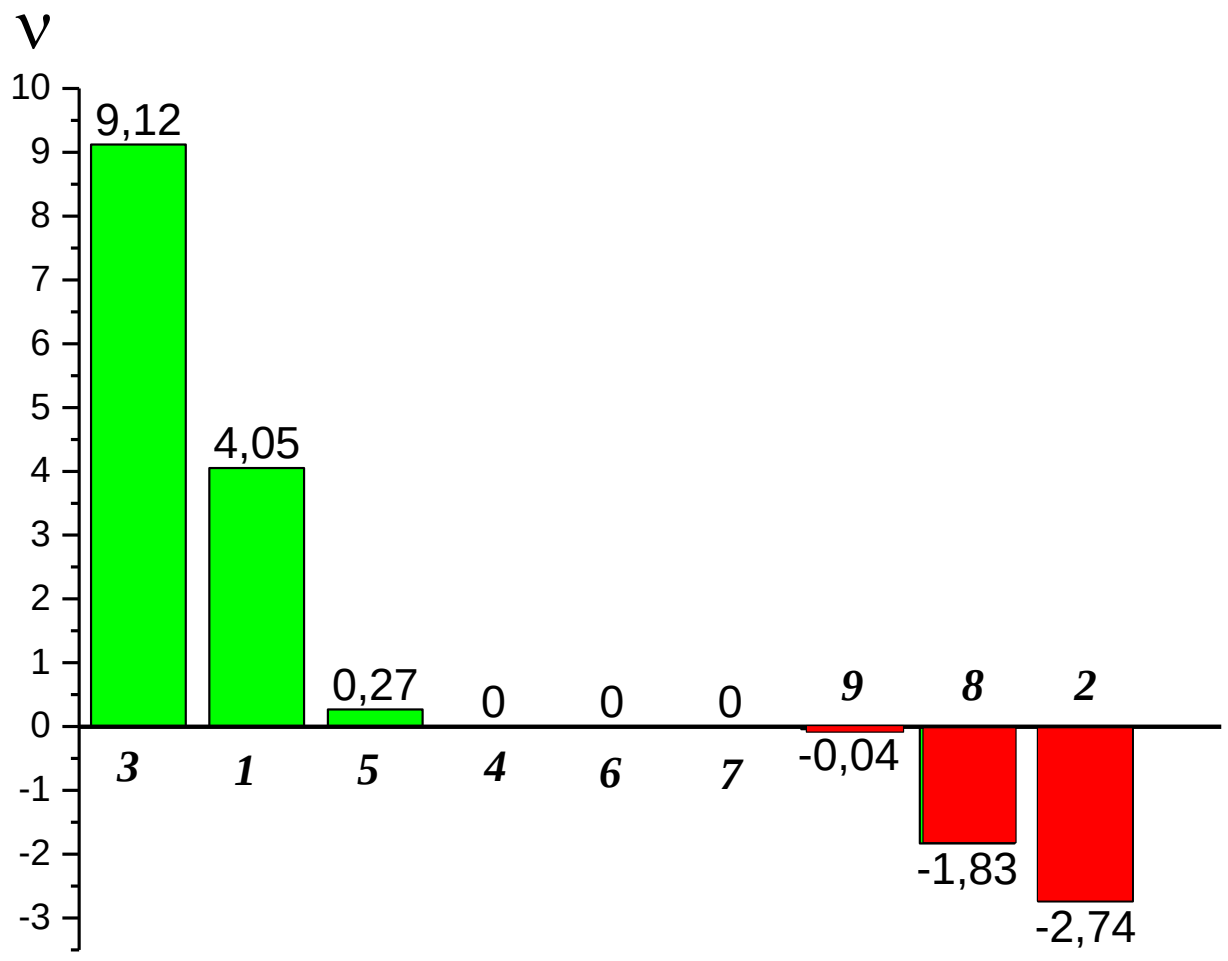

(d)

$$\nu = \begin{pmatrix} 4.05 \\ -2.74 \\ 9.12 \\ 0 \\ 0.27 \\ 0 \\ 0 \\ -1.83 \\ -0.04 \end{pmatrix}$$

Fig. 6. "Pressure" - $\psi$ and "influence" $\nu$ for the cognitive map shown in Fig. 5. (a). - ranking of the location of the number of nodes (c), b and d - vector $\psi$ and $\nu$, respectively.

As can be seen from Fig. 6 the "influence" factor in the general network has the "population" factor, followed by "the length of a travel» and behind them the "cost of the travel" with a large margin.

As expected, the factor experiencing the biggest "pressure" is "the volume of exhaust gases emissions into the atmosphere", followed by "accidents" (who would have thought?) And "fuel consumption".

Weight of the nodes is calculated in the approach of the impulse method. Calculations fulfilled by the given lines on the influence of one node on another (the adjacency matrix $W$ (Fig. 5)). For comparison with the $K$-method proposed here, one can introduce characteristics $\psi^{imp}$ and $\nu^{imp}$ analogously to pressure and influence, for the impulse method (see apendix A).

The results of the calculation of $\psi$ and $\nu$ for the cognitive map (Fig. 5) are shown in Fig. 7.

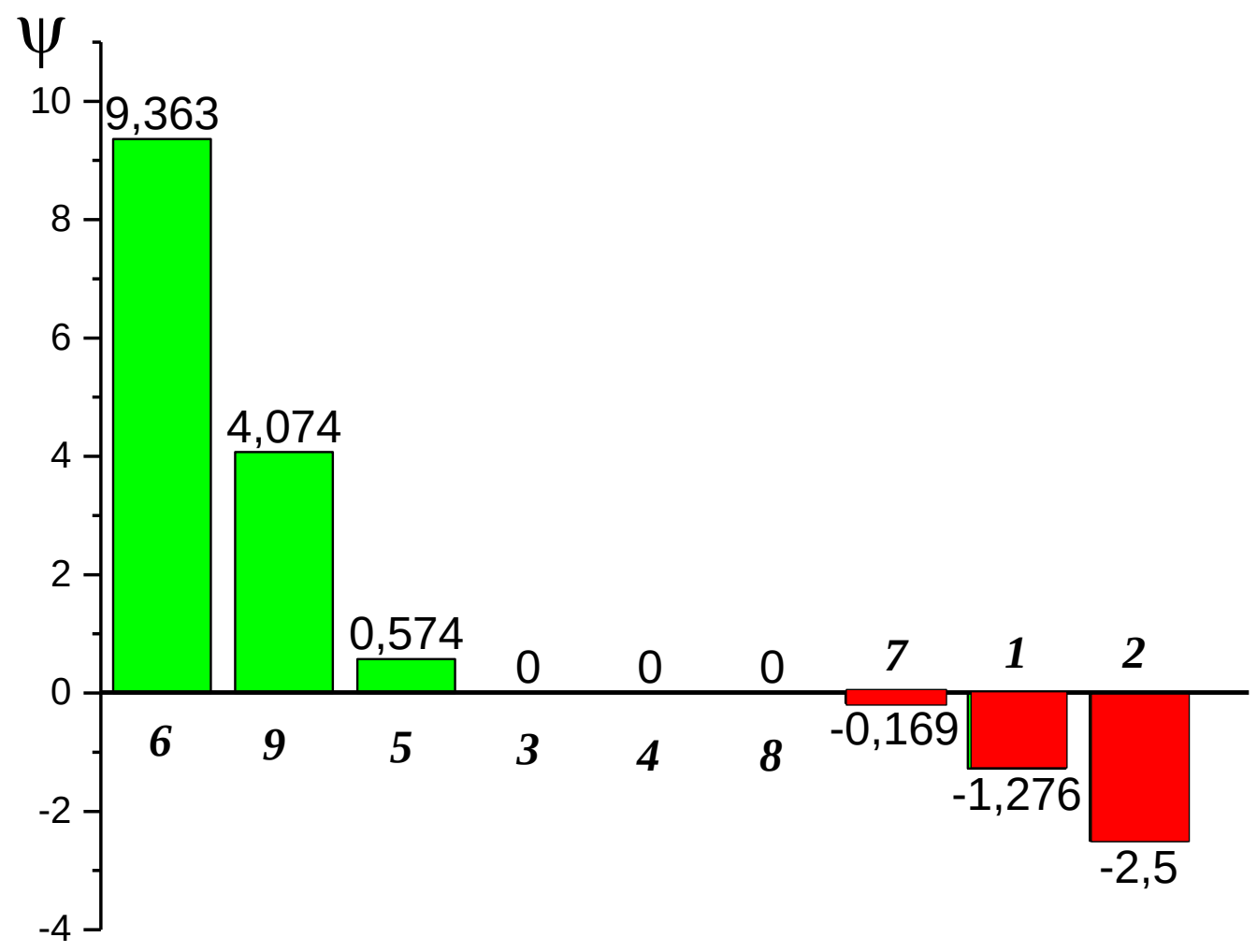

(a)

$$\psi = \begin{pmatrix} -1.276 \\ -2,5 \\ 0 \\ 0 \\ 0.574 \\ 9.363 \\ -0.169 \\ 0 \\ 4.074 \end{pmatrix}$$ (b)

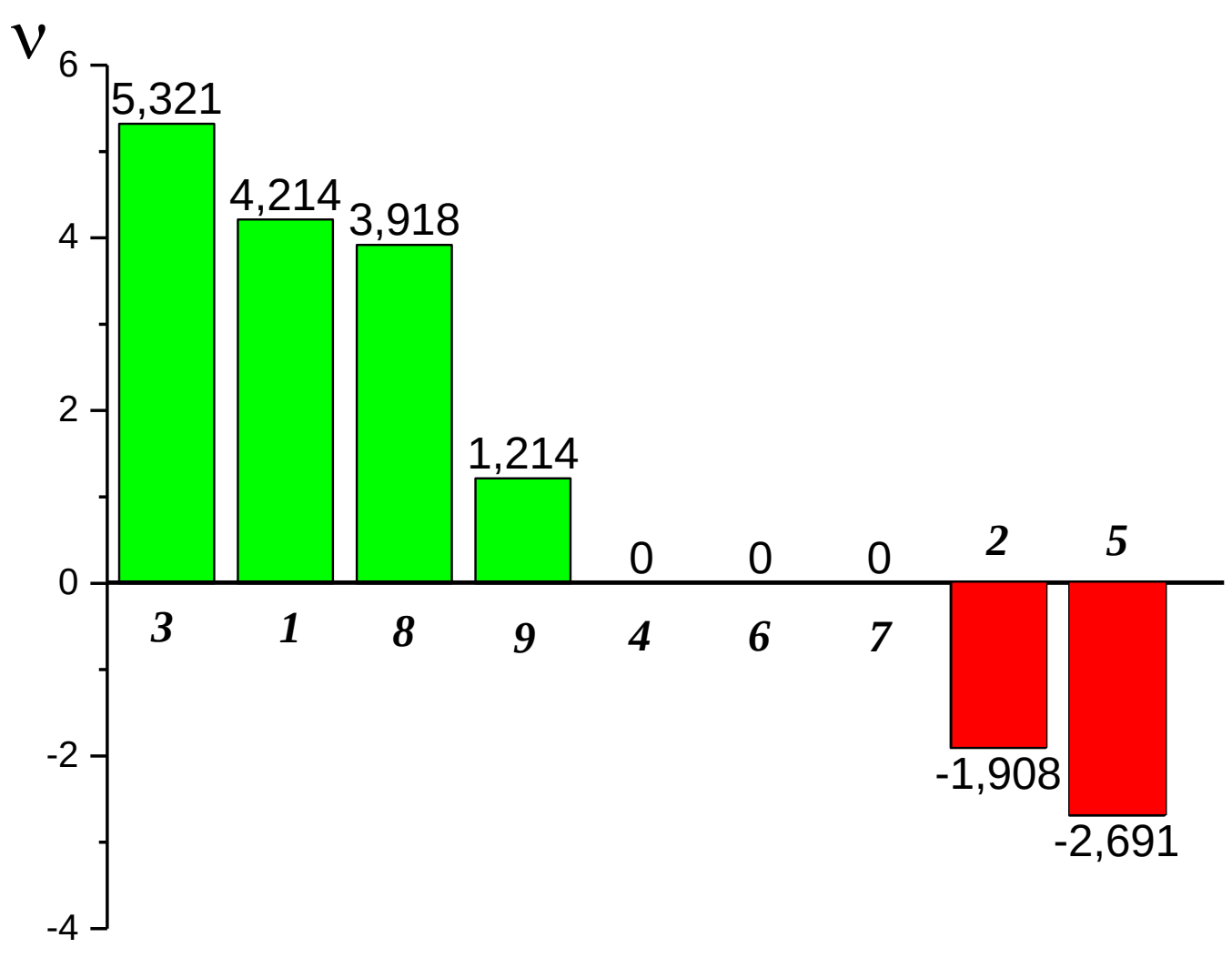

(c)

$$\nu = \begin{pmatrix} 4.214 \\ -2.691 \\ 5.321 \\ 0 \\ -1.908 \\ 0 \\ 0 \\ 3.918 \\ 1.214 \end{pmatrix}$$ (d)

Fig. 7. "Pressure" and "influence" in the impulse method, ranking of the location of the number of nodes (a, c), vector $\psi^{imp}$ (b) and vector $\nu^{imp}$ (d).

Table 2. Comparison of the ranking of coefficients (nods) fulfilled by K- and impulse methods, respectfully.

| | | | | | | | | | |
|---|---|---|---|---|---|---|---|---|---|
| $\psi$ | 8 | 9 | 5 | 6 | 4 | 1 | 2 | 7 | 3 |
| $\psi^{imp}$ | 7 | 9 | 4 | 5 | 3 | 1 | 8 | 6 | 2 |
| $\nu$ | 2 | 9 | 1 | 4 | 3 | 5 | 6 | 8 | 7 |
| $\nu^{imp}$ | 2 | 9 | 1 | 5 | 8 | 6 | 7 | 3 | 4 |

The ranking of nodes by a numerical value by the K- and the impulse methods (Table 1) is similar to each other (excepting a few nodes). Namely, for $\nu$ and $\nu^{imp}$ with large positive values of 2, 9 and 1. A high positive value in the case of application of K-method and close to zero in the using the impulse method. To characterize the "influence" of such, the different nodes are two to five and eight in rank.

Despite some similarity of the results of nods ranking by the K-method and the impulse method, it should be noted that the impulse method does not satisfy the requirements formed above. In particular, when the adjacency matrix is changed to the same value, the ranking of the coefficients (nods) can be changed. And in some cases this leads to the fact that calculations by the impulse method generally cease to converge.

## Conclusions

The article suggests new pair characteristics of complex network nodes and the method of their calculation (K-method), which can be used to calculate the mutual influence of concepts in cognitive maps. Due to simplifications in the proposed method, it becomes possible to use the simplified calculation mechanism, similar to one, is applicable in electrical engineering. Two new weight characteristics of the concept nodes are proposed, namely pressure and influence as results of accounting for the network structure and "collective interaction".

It should be noted that the proposed algorithm is devoid of the main problems and contradictions inherent in the widely used impulse method [9] [12], namely:

1. Divergence of the values of the connection weights when the series of degrees of the adjacency matrix $W^k$ diverges.

2. The increase in the elements of the adjacency matrix $W$ by the same amount not only changes the magnitude of the components of the resulting vector, but also changes their order in the ranking.

At the same time, the results obtained with the help of the K-method for a real network are correlating well enough with the results obtained by the impulse method. The advantages of the proposed K-method include its computational simplicity (in comparison with other known algorithms) comparable with the task of enumerating subgraphs for sparse networks of relatively small size (in practice - several hundred nodes).

## Appendix A

According to the impulse method [8], each component (node) receives initial value $V_i(init)$. Thus, a vector of $\vec{V}_i(init)$ can be formed. Next, a rule for determining $\vec{V}(n)$ is introduced at each next moment of discrete "time" $n = 0,1,...$

$$\vec{V}(n+1) = \vec{V}(n) + \hat{W}\vec{p}(0),\ n = 0,1,... \tag{A1}$$

where $\vec{V}(n)$ is the column vector of the nodes of the cognitive map, $\hat{W}$ is the contiguity matrix of the cognitive map,

$$\vec{p}(n) = \vec{V}(n) - \vec{V}(n-1),\ n = 1,2,... \tag{A2}$$

and at some moment of time $n = 0$ are considered given $\vec{p}(0)$ and $\vec{V}(init)$:

$$\vec{V}(0) = \vec{V}(init) + \vec{p}(0) \tag{A3}$$

Thus, the equation for the determination $\vec{V}(n)$ is given as follows [8]:

$$\vec{V}(n) = \vec{V}(init) + \sum_{k=0}^{n} \hat{W}^k \cdot p(0) \tag{A4}$$

In the case when the series (A4) converges can be expressed in terms of the matrix inversed to $\hat{I} - \hat{W}$, where is $\hat{I}$ the unity matrix.

$$\vec{V}(\infty) = \vec{V}(init) + (1 - \hat{W})^{-1} \cdot \vec{p}(0)$$

Later on we denote $\vec{V}(\infty)$ as $\vec{V}$, and $(1-\hat{W})^{-1} = \hat{\Omega}$ (A5)

considering, $\vec{V}(init)$ =0, thus

$$\vec{V} = \hat{\Omega} \cdot \vec{p}(0) \tag{A6}$$

Analog of pressure concept $\alpha$ is $\psi_\alpha$ . In impulse method may value of $\alpha$ is component of vector $\Omega p(0)$, where $p(0)$ has zero $\beta$ component and the rest components are equal to 1. So, for a cognitive map with three concepts $\psi_2^{imp}$ is defined as: значение альфа может быть компонентом вектора

$$\psi_2^{imp} = \left[ \begin{pmatrix} \Omega_{11} & \Omega_{12} & \Omega_{13} \\ \Omega_{21} & \Omega_{22} & \Omega_{23} \\ \Omega_{31} & \Omega_{32} & \Omega_{33} \end{pmatrix} \begin{pmatrix} 1 \\ 0 \\ 1 \end{pmatrix} \right] = \Omega_{21} + \Omega_{23} \quad \text{(A7)}$$

Thereby, at the initial moment of time unity impulses on all nodes are set, except the second one, then values $\psi_2^{imp}$ on concept 2 with $n \to \infty$ can be calculated. Thus, one may represent $\psi_2^{imp}$ as total action of all concepts on the second concept.

For arbitrary component $\beta$ value $\psi_\beta^{imp}$ may be defined as follows:

$$\psi_\beta^{imp} = \sum_k \Omega_{k\beta} - \Omega_{\beta\beta} \quad \text{(A8)}$$

Similarly, for impulse analog of consequence $\nu_\alpha^{imp}$, following relation takes place:

$$\nu_\alpha^{imp} = \sum_k \Omega_{\alpha k} - \Omega_{\alpha\alpha} \quad \text{(A9).}$$